\def\year{2021}\relax
\documentclass[letterpaper]{article} % DO NOT CHANGE THIS
\pdfoutput=1
\usepackage{aaai21}  % DO NOT CHANGE THIS
\usepackage{times}  % DO NOT CHANGE THIS
\usepackage{helvet} % DO NOT CHANGE THIS
\usepackage{courier}  % DO NOT CHANGE THIS
\usepackage[hyphens]{url}  % DO NOT CHANGE THIS
\usepackage{graphicx} % DO NOT CHANGE THIS
\usepackage[algo2e]{algorithm2e}
\makeatletter
\newcommand{\algorithmstyle}[1]{\renewcommand{\algocf@style}{#1}}
\makeatother
\usepackage{natbib}  % DO NOT CHANGE THIS AND DO NOT ADD ANY OPTIONS TO IT
\usepackage{caption} % DO NOT CHANGE THIS AND DO NOT ADD ANY OPTIONS TO IT
\usepackage{color}

\title{Evaluating Generalization and Transfer Capacity of Multi-Agent Reinforcement Learning Across Variable Number of Agents}
\author {
    Bengisu Guresti,\textsuperscript{\rm 1}
    Nazim Kemal Ure \textsuperscript{\rm 1} \\
}
\affiliations {
    \textsuperscript{\rm 1} Istanbul Technical University \\
    guresti15@itu.edu.tr, ure@itu.edu.tr
}

\begin{document}

\maketitle

\begin{abstract}
Multi-agent Reinforcement Learning (MARL) problems often require cooperation among agents in order to solve a task. Centralization and decentralization are two approaches used for cooperation in MARL. While fully decentralized methods are prone to converge to suboptimal solutions due to partial observability and nonstationarity, the methods involving centralization suffer from scalability limitations and lazy agent problem. Centralized training decentralized execution paradigm brings out the best of these two approaches; however, centralized training still has an upper limit of scalability not only for acquired coordination performance but also for model size and training time. In this work, we adopt the centralized training with decentralized execution paradigm and investigate the generalization and transfer capacity of the trained models across variable number of agents. This capacity is assessed by training variable number of agents in a specific MARL problem and then performing greedy evaluations with variable number of agents for each training configuration. Thus, we analyze the evaluation performance for each combination of agent count for training versus evaluation.  We perform experimental evaluations on predator prey and traffic junction environments and demonstrate that it is possible to obtain similar or higher evaluation performance by training with less agents. We conclude that optimal number of agents to perform training may differ from the target number of agents and argue that transfer across large number of agents can be a more efficient solution to scaling up than directly increasing number of agents during training.
\end{abstract}

\section{Introduction}
\noindent
%Agents need to cooperate to achieve a common goal in multi-agent systems. 
A significant amount of problems in Multi-agent Reinforcement Learning (MARL) require agents to achieve a common goal, which makes learning to cooperate crucial. Nevertheless, as the number agents that need to cooperate increases, the difficulty of achieving optimal cooperation also increases. In order to create large-scale MARL applications, it is critical to find high performance solutions that facilitate scalability with low computational costs. As a step towards this goal, we pose the question: can learned policies be transferred across systems with higher number of agents without any extra training necessary and loss of performance?

Centralization and decentralization are two main paradigms in cooperative MARL.
Centralization aims to reach cooperation naturally by transforming the multi-agent problem into a single agent problem by aggregating the local observations of agents to form a global state and inferring actions of agents through the global state. However, besides its inherent problems, full centralization with conventional function approximators does not allow execution on systems with different number of agents than the number of agents used in training, which we call transfer across variable number of agents. Nevertheless, use of graph convolutional networks \cite{kipf2017semisupervised}, \cite{veli} in a centralized training centralized execution paradigm as in \cite{li2020deep} allows transfer to systems with variable number of agents.

Centralized training decentralized execution paradigm makes transfer across variable number of agents straightforward, since the obtained policy infers actions from local observations and shares parameters with every agent in the system. Apart from that advantage, the overhead from increasing number of agents during execution is relatively small with respect to centralized execution. Therefore, we adopt centralized training decentralized execution as our main approach. Furthermore, we base our approach on graph convolutional MARL methods such as in \cite{jiang2020graph}, \cite{li2020deep} that represent the locality and strength of agent interactions and form an implicit coordination graph.

In order to evaluate the generalization and transfer capacity across variable number of agents, we perform training in predator prey and traffic junction environments with varying number of agents starting from two-three agents to the number of agents that the environment's capacity allows. We then perform greedy evaluations on the trained models with varying number of agents. We analyze the performance of evaluation results in terms of generalization and transfer capacity for each combination of agent count for training versus evaluation. The experiment results demonstrate the possibility of obtaining similar or higher evaluation performance by training with less agents. We argue that training with a smaller number of agents and then transferring the model to high-scale configurations can be a more efficient solution than training with the high-scale configurations for agent count while preserving performance.

\section{Related Work}
\noindent
Centralization and decentralization are central approaches in cooperative MARL.
Centralization aims to reach cooperation naturally by transforming the multi-agent problem into a single agent problem. However, \cite{DBLP:journals/corr/SunehagLGCZJLSL17} demonstrate that centralization leads to inefficient policies due to the lazy agent problem where some agents refrain from learning while an agent learns successfully because those agents' exploratory behaviour would damage the agent's learning performance. Furthermore, achieving centralization is not practically possible for some tasks due to impossibility or impracticability of being informed of other agents' observations. The alternative approach, and the mandatory approach when centralization is not possible, is decentralization where each agent is an independent learner. However, \cite{DBLP:journals/corr/SunehagLGCZJLSL17} assert that nonstationarity introduces spurious reward signals that an agent can not determine if the signal is the outcome of its own action or other agents' actions, thus leading to failure. In order to mitigate these drawbacks, it is a common approach to use the centralized training decentralized execution paradigm. Counterfactual Multi-Agent Policy Gradients (COMA) is a classic example of this paradigm which is an actor critic algorithm with a centralized actor and a decentralized critic \cite{foerster2017counterfactual}. We also use an actor-critic algorithm, Proximal Policy Optimization (PPO), as our base algorithm. PPO \cite{schulman2017proximal} is a policy gradient method that optimizes a surrogate objective function that enables the algorithm to learn while limiting the extent the policy may change in each iteration.

Instead of adopting either centralized or decentralized approach, \cite{NIPS2001_1941} propose formulating the cooperation problem by forming coordination graph and transforming this graph into a Dynamic Bayesian Network (DBN) for factorized representation, which allows factoring value functions in order to enable agents to coordinate by message passing and solved by linear programming. \cite{bohmer2020deep} applies coordination graph approach to deep neural networks and approximate pay-off functions using them while maximizing value function by message passing. 

Considering the convenience of representing multi-agent dynamics as a graph, processing graph structured data using deep neural networks is of importance. \cite{kipf2017semisupervised} propose a layer-wise propagation rule for deep neural networks processing graph structured data. \cite{veli} introduce Graph Attention Networks which improve upon previous methods by using a masked self-attention layer that weights the impact of each neighbor accordingly during aggregation. \cite{jiang2020graph} propose using graph convolution with relation kernels in MARL to capture agent interplay that adapts to underlying dynamic graph of the environment in order to promote cooperation. \cite{li2020deep} suggest using self attention to obtain the coordination graph structure once and then using it for graph convolution in each pass to form an implicit deep coordination graph.

Our work adopts centralized training decentralized execution paradigm that uses PPO. We use graph convolution with self attention to process graph structured data and encourage cooperation. Although the aforementioned related work also aims to promote cooperation, they mainly focus on achieving it for a fixed number of agents. Our contribution is checking the limits of cooperation that is learned for a fixed number of agents across different number of agents, and demonstrating that increasing agent count beyond a threshold in training is not necessary for achieving cooperation.

\section{Methodology}

\subsection{Algorithm and Network Architecture}

We adopt the Decentralized Partially Observable Markov Decision Process (Dec-POMDP) framework as formulated in \cite{Oliehoek2012} for this MARL problem. We use PPO with clipped surrogate objective as used and parametrized in \cite{li2020deep}. We use a simple Multi-Layer Perceptron (MLP) that takes single local observations and outputs action probabilities for each agent. %The network architecture for actor is given in Figure \ref{fig:figure_actor}.

% \begin{figure}
%     \centering
%     \includegraphics[width=0.2\textwidth]{new_figure3.png}
%     \caption{Actor Network - $o_i$ denotes observation of agent $i$, $Pr(act_i)$ denotes probability distribution of action $act_i$ taken by agent $i$ }
%     \label{fig:figure_actor}
% \end{figure}

\begin{figure}
    \centering
    \includegraphics[width=0.45\textwidth]{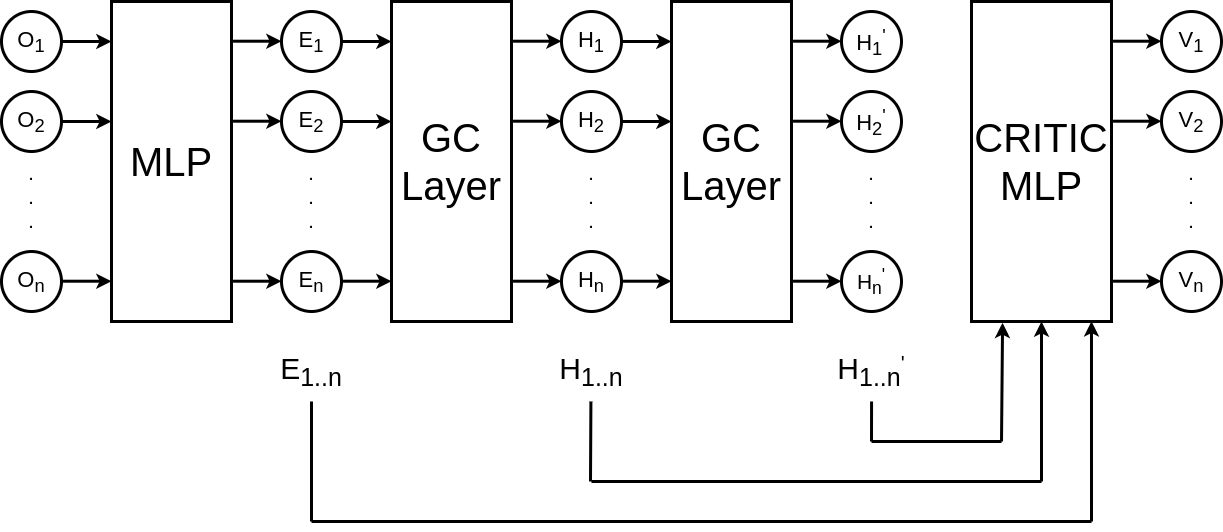}
    \caption{Critic Network - $o_i$ denotes observation of agent $i$, $E_i$ denotes embedding of the observation of agent $i$, $H_i$ denotes output of agent $i$ from the first graph convolution, $H'_i$ denotes output of agent $i$ from the second graph convolution, $v_i$ denotes the corresponding value of agent $i$}
    \label{fig:figure_critic}
\end{figure}

The network architecture of centralized critic is given in Figure \ref{fig:figure_critic}. The architecture consists of one MLP for extracting embeddings of decentralized observations. Then these embeddings are forwarded to two layers of graph convolutional layers with self attention which use message passing to aggregate these local observations properly. Then both embeddings and outputs of graph convolutional layers are forwarded to Critic MLP module which first aggregates information using sum operation and forwards it through MLP.

\begin{figure}
    \centering
    \includegraphics[width=0.5\textwidth]{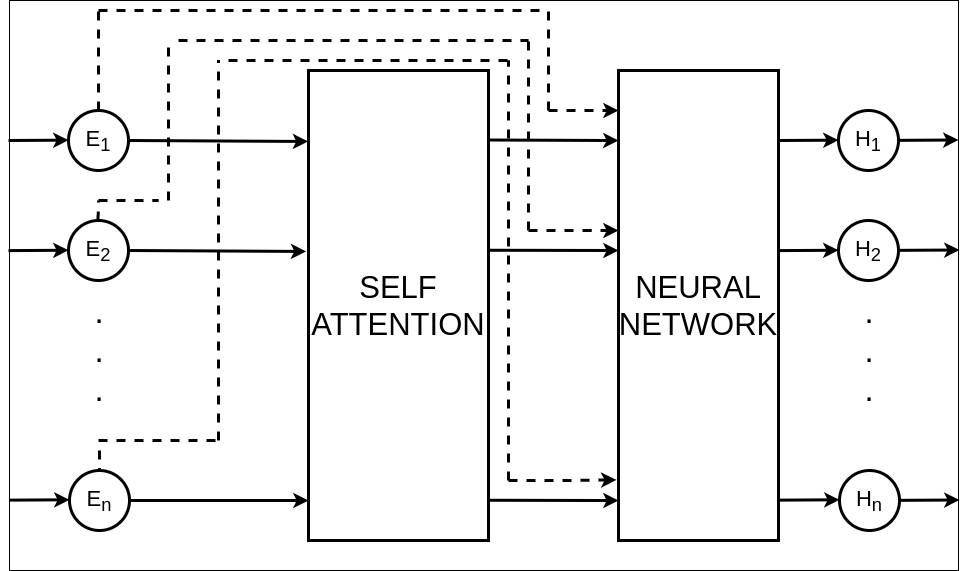}
    \caption{Graph Convolutional Layer - $E_i$ denotes the input of agent $i$ to the graph convolution, $H_i$ denotes the output of agent from the graph convolution}
    \label{fig:figure_gcn}
\end{figure}

The network architecture of graph convolutional layer is provided in Figure \ref{fig:figure_gcn}. This layer passes each of its inputs through self attention, then it forwards attention outputs and residuals to a one layer neural network which produces graph convolutional layer outputs. The used self attention formulation and graph convolution outputs are inspired from \cite{jiang2020graph} and formulated in Equation \ref{eq:att} and Equation \ref{eq:gconv}. In these equations, h denotes input to the graph convolutional layer, $h^{'}$ denotes output of the graph convolutional layer, $d_{k}$ denotes the scaling factor, $\sigma$ denotes one layer feed forward network, and $W_{Q}$, $W_{K}$, $W_{V}$ denote weight matrices of query, key, and value vectors.

\begin{equation}
\textnormal{Self-Attention}(h)=\textnormal{softmax}\left(\frac{\mathbf{W}_{Q} h (\mathbf{W}_{K} h)^{T}}{\sqrt{d_{k}}}\right)\mathbf{W}_{V} h
\label{eq:att}
\end{equation}

\begin{equation}
h^{\prime}=\sigma\left(\textnormal{ concatenate }\left[\textnormal{Self-Attention}(h), h\right]\right)
\label{eq:gconv}
\end{equation}

\subsection{Generalization and Transfer Capacity Evaluation}
\noindent
In order to evaluate the generalization and transfer capacity of the agents, we determine a range of agent count that starts from 2-3 and reaches to the capacity of the environment in question. Since we use decentralized execution that shares the actor with all agents, transfer is achieved simply by using this actor for all the agents with their local observations. We perform training in our chosen MARL environments for the whole range of agent count with 3 different seeds. We then determine a range of agent count for evaluation. The evaluation agent count range coincides with the training agent count range except for some additional sample points we may choose from inside the training range. The reason for choosing additional sample points from inside the training range is to make sure that we can answer the question: would training a model with higher agent count give superior results? Each evaluation process takes 100 evaluations averaged over with 3 different seeds in the environment, thus every combination contains 9 different evaluation results. At the end, evaluations of models with the same agent count for training and same agent count for evaluation are averaged and their standard deviations are calculated. For analysis, evaluation results are grouped according to the agent count during evaluation, in order to answer the question: is training $n$ agents the optimal choice for applications where $n$ or higher number of agents will be necessary, or is it possible to train with fewer number of agents and transfer the model to applications with higher number of agents necessary with similar or better performance?
\newline

\algorithmstyle{plain}
\begin{algorithm2e}
\caption{Transfer Capacity Evaluation}
 \label{algo:tce}
\SetKwInOut{Input}{input}
\SetKwInOut{Output}{output}
\Input{number of training epochs ($ep$), train agent count ($tac$), eval agent count ($eac$), train seeds, evaluation seeds}
\Output{$A$}\tcp{A: 2d array with shape $eac, tac$}
\SetAlgoLined
   \For{each $n$ in $tac$}{
       \For{each $seed$ in train seeds}{
            \For{$epoch\gets1$ to \KwTo $ep$}{
                Train $agent_n$ for $tac$ $n$
            }
            save $agent_n$ with train seed: $seed$ and $tac$: $n$ 
        }
    }
   \For{each $n_{eval}$ in $eac$}{
       \For{each $n_{train}$ $tac$}{
            $result_{eval}\gets0$
            
            \For{each $agent$ in saved agents with train agent count = $n_{train}$}{
                \For {each $seed$ in evaluation seeds}{
                    $result_{avg} \gets $ evaluate $100$ times
                    
                    $result_{eval}\gets result_{eval} + result_{avg}$
                }
            }
            $result_{eval} \gets result_{eval}$ / (\# of train seeds $\times$ \# of evaluation seeds)\\
            $A[n_{eval}][n_{train}] \gets result_{eval}$ 
        }
    }
\end{algorithm2e}

\section{Experiments}
\noindent

\subsection{Predator Prey}
Predator Prey environment consists of preys and predators where predators get rewarded by hunting preys, which move by hard coded action descriptions and random moves as described in \cite{li2020deep}. As applied by \cite{li2020deep}, we also penalize single agent capture attempts by -0.5 penalty in order to compel predators to collaborate.  
We use predator prey environment with a grid size of 20 $\times$ 20 in order to create capacity for 80 preys and 80 predators. We determine the training range and evaluation range of agent count as: 2, 5, 10, 20, 50, 80. The mean evaluation rewards are provided in Table \ref{tab:predator1} and Table \ref{tab:predator2}.

\begin{table}[]
\centering
{\small
\begin{tabular}{llll}

\hline
   &$ 2          $&$ 5         $&$ 10$\\ \hline
$2  $&$ -2.01\pm1.56  $&$ 30.30\pm1.95 $&$ 86.48\pm2.69 $\\
$5  $&$ +4.62\pm1.46   $&$ 41.98\pm0.77 $&$ 94.60\pm0.78 $\\
$10 $&$ -2.93\pm1.10  $&$ 39.71\pm1.56 $&$ 95.35\pm0.21 $\\ 
$20 $&$ -11.75\pm2.09 $&$ 22.05\pm6.95 $&$ 84.60\pm7.86 $\\
$50 $&$ -16.52\pm1.68 $&$ 6.68\pm4.03  $&$ 63.39\pm4.61 $\\
$80 $&$ -13.96\pm1.00 $&$ 11.17\pm4.29 $&$ 62.57\pm7.94 $\\

\hline

\end{tabular}
}
\caption{Mean of Total Rewards for Predator Prey \newline (Columns denote the number of agents in evaluation while rows the denote number of agents in training.) } \label{tab:predator1}

\end{table}
\begin{table}[]
\centering
{\small
\begin{tabular}{llll}

\hline
   &$ 20         $&$ 50         $&$ 80 $        \\ \hline
$2  $&$ 192.38\pm1.88 $&$ 496.58\pm0.56 $&$ 797.30\pm0.59 $\\
$5  $&$ 196.37\pm0.58 $&$ 497.96\pm0.40 $&$ 798.07\pm0.82 $\\
$10 $&$ 196.95\pm0.07 $&$ 498.18\pm0.37 $&$ 798.28\pm0.76 $\\ 
$20 $&$ 194.08\pm2.48 $&$ 496.80\pm0.47 $&$ 794.74\pm2.30 $\\
$50 $&$ 182.00\pm2.86 $&$ 494.68\pm1.00 $&$ 795.90\pm1.45 $\\
$80 $&$ 168.27\pm8.47 $&$ 484.42\pm4.33 $&$ 789.59\pm2.09 $\\
\hline

\end{tabular}
}

\caption{Mean of Total Rewards for Predator Prey \newline (Columns denote the number of agents in evaluation while rows the denote number of agents in training.) } \label{tab:predator2}

\end{table}

%Results of the experimental evaluation in predator prey environment, as provided in Table \ref{tab:predator1}, Table \ref{tab:predator2}, Figure \ref{fig:figure_pp5} and Figure \ref{fig:figure_pp50},

\begin{figure}
    \centering
    \includegraphics[width=0.4\textwidth]{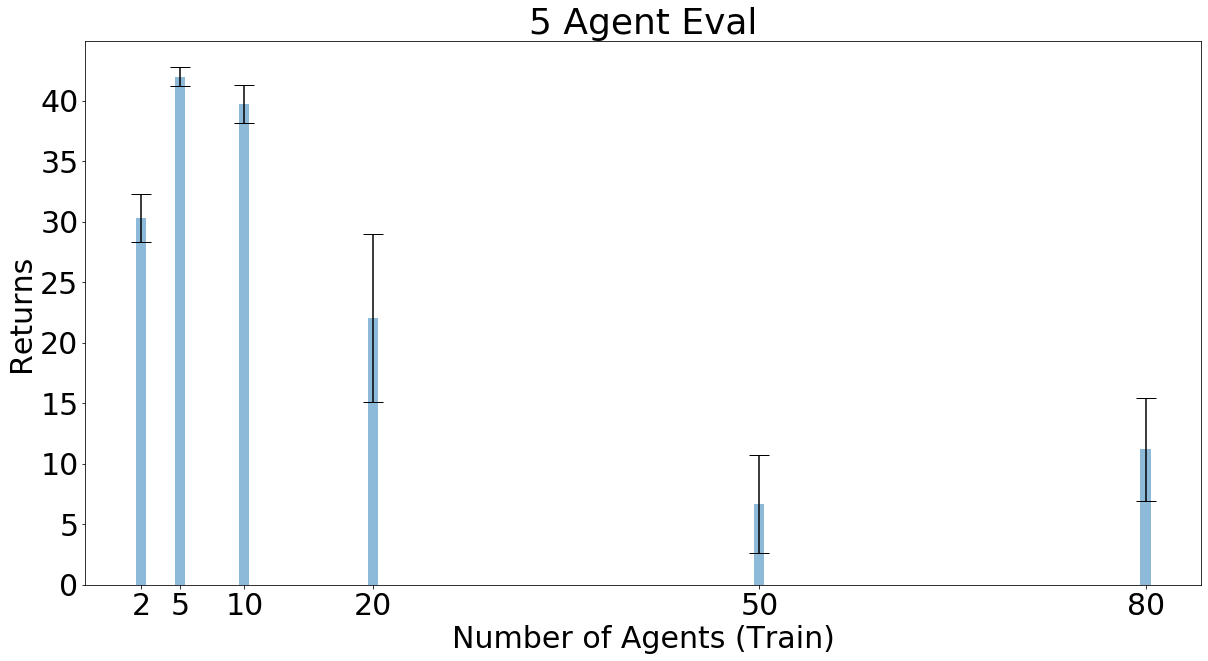}
    \caption{5 Agents Evaluation in Predator Prey}
    \label{fig:figure_pp5}
\end{figure}

Results of the experimental evaluation in predator prey environment, as provided in Table \ref{tab:predator1}, Table \ref{tab:predator2}, and Figure \ref{fig:figure_pp5}, demonstrate that training in few number of agents such as 2-5 can get evaluation results that surpass the evaluation results of models that are trained with large number of agents such as 50-80. It can be deduced that models trained with few number of agents have high generalization and transfer capacity for execution with high number of agents. However, our analysis shows us that the reverse is not true. The models that are trained with large number of agents such as 50-80 have very low returns for the evaluation cases where there are 2-5 agents in the environment. Hence, it can be inferred from the results that for a high performance application of predator prey environment, choosing number of agents to train from the range [5, 10] would be the better choice.

% \begin{figure}
%     \centering
%     \includegraphics[width=0.4\textwidth]{pp_50_agent.png}
%     \caption{50 Agents Evaluation in Predator Prey}
%     \label{fig:figure_pp50}
% \end{figure}

\subsection{Traffic Junction}
Traffic Junction environment consists of predetermined routes and junctions where cars need to reach their destination point without collision as described in \cite{li2020deep}. We use traffic junction environment with the difficulty 'hard'. Other parameters of the environment such as dimension, maximum agent add rate and minimum add agent rate are chosen compatibly with the environment difficulty according to the configurations proposed by \cite{singh2018learning}. Because of environment dimension and agent add rate setting, the capacity of the environment allows for approximately 20 agents in a single time step. Thus, we determine the training range and evaluation range of agent count as: 3, 5, 10, 15, 20. The mean evaluation success rates are provided in Table \ref{tab:traffic1}.

\begin{table}[]
\centering
\setlength\tabcolsep{2.5pt}
{\small
\begin{tabular}{llllll}

\hline
   & $ 3           $ & $ 5           $ & $ 10          $ & $  15          $ & $ 20 $    \\ \hline
$3  $ & $ 0.99\pm0 $ & $ 0.92\pm0.04 $ & $ 0.56\pm0.10 $ & $ 0.26\pm0.14  $ & $ 0.23\pm0.15$\\
$5 $ & $ 0.99\pm0 $ & $ 0.97\pm0.01 $ & $ 0.77\pm0.09 $ & $ 0.58\pm0.16  $ & $ 0.58\pm0.18$\\
$10 $ & $ 1.00\pm0 $ & $ 0.99\pm0.00 $ & $ 0.95\pm0.01 $ & $ 0.84\pm0.07  $ & $ 0.73\pm0.09$\\
$15 $ & $ 1.00\pm0 $ & $ 0.99\pm0.01 $ & $ 0.94\pm0.01 $ & $ 0.85\pm0.03  $ & $ 0.79\pm0.05$\\
$20 $ & $ 0.99\pm0 $ & $ 0.99\pm0.00 $ & $ 0.90\pm0.02 $ & $ 0.83\pm0.04  $ & $ 0.79\pm0.03$\\

\hline
\end{tabular}
}

\caption{Mean of Success Rates for Traffic Junction \newline (Columns denote the number of agents in evaluation while rows denote the number of agents in training.) } \label{tab:traffic1}

\end{table}

\begin{figure}
    \centering
    \includegraphics[width=0.4\textwidth]{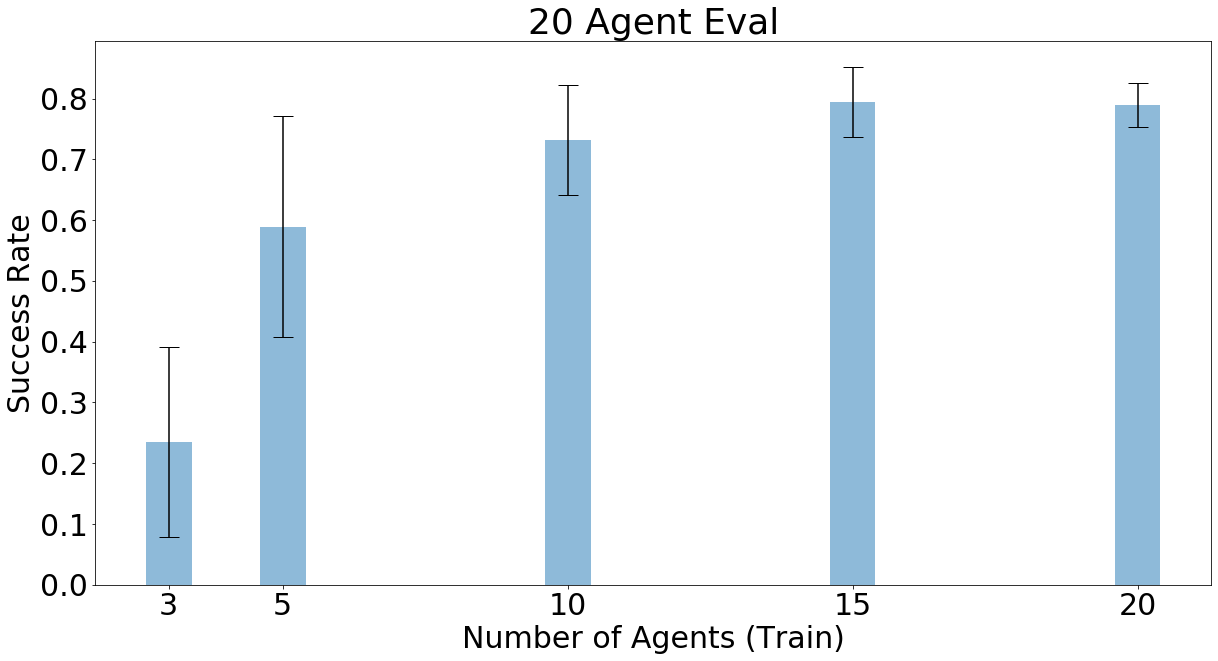}
    \caption{20 Agents Evaluation in Traffic Junction}
    \label{fig:figure_tr20}
\end{figure}

Results of the experimental evaluation in traffic junction environment, as provided in Table \ref{tab:traffic1} and Figure \ref{fig:figure_tr20}, demonstrate that training with few number of agents such as 3-5 gets evaluation results with much lower success rate compared to the evaluation results of models that are trained with large number of agents such as 15-20. It can be deduced that models trained with few number of agents can not sufficiently transfer for execution with high number of agents. The models that are trained with 15-20 agents have very high success rate for the evaluation cases where there are 3-5 agents in the environment. Nevertheless, the evaluation results with 15 and 20 agents demonstrate that training with 15 agents gives better results than training with 20 agents. Thus, we infer that an environment can have a sweet spot for the number of agents to train. We also infer that environment dynamics play a key role in the generalization and transfer capacity of training.

\section{Discussion}

\noindent
In this work, we adopted the centralized training decentralized execution paradigm and investigated the generalization and transfer capacity of the trained models across variable number of agents. We assessed it by training variable number of agents in a specific MARL problem and then performing greedy evaluations with variable number of agents for each training configuration. We deduced that an environment can have a sweet spot for the number of agents to train in terms of evaluation performance and that environment dynamics play a key role in the generalization and transfer capacity of training. We saw that training with fewer number of agents can be a more efficient option for execution in large number of agents. We conclude that optimal number of agents to perform training may differ from the target number of agents and put forward that transfer across large number of agents can be a more efficient solution to scaling up than directly increasing number of agents during training.

\section{Acknowledgments}
This work is supported by the Scientific Research Project
Unit (BAP) of Istanbul Technical University, Project Number: MOA-2019-42321

\noindent
\bibliography{references}

\begin{thebibliography}{11}
\providecommand{\natexlab}[1]{#1}
\providecommand{\url}[1]{\texttt{#1}}
\providecommand{\urlprefix}{URL }
\expandafter\ifx\csname urlstyle\endcsname\relax
  \providecommand{\doi}[1]{doi:\discretionary{}{}{}#1}\else
  \providecommand{\doi}{doi:\discretionary{}{}{}\begingroup
  \urlstyle{rm}\Url}\fi

\bibitem[{Böhmer, Kurin, and Whiteson(2020)}]{bohmer2020deep}
Böhmer, W.; Kurin, V.; and Whiteson, S. 2020.
\newblock Deep Coordination Graphs.

\bibitem[{Foerster et~al.(2017)Foerster, Farquhar, Afouras, Nardelli, and
  Whiteson}]{foerster2017counterfactual}
Foerster, J.; Farquhar, G.; Afouras, T.; Nardelli, N.; and Whiteson, S. 2017.
\newblock Counterfactual Multi-Agent Policy Gradients.

\bibitem[{Guestrin, Koller, and Parr(2002)}]{NIPS2001_1941}
Guestrin, C.; Koller, D.; and Parr, R. 2002.
\newblock Multiagent Planning with Factored MDPs.
\newblock In Dietterich, T.~G.; Becker, S.; and Ghahramani, Z., eds.,
  \emph{Advances in Neural Information Processing Systems 14}, 1523--1530. MIT
  Press.
\newblock
  \urlprefix\url{http://papers.nips.cc/paper/1941-multiagent-planning-with-factored-mdps.pdf}.

\bibitem[{Jiang et~al.(2020)Jiang, Dun, Huang, and Lu}]{jiang2020graph}
Jiang, J.; Dun, C.; Huang, T.; and Lu, Z. 2020.
\newblock Graph Convolutional Reinforcement Learning.

\bibitem[{Kipf and Welling(2017)}]{kipf2017semisupervised}
Kipf, T.~N.; and Welling, M. 2017.
\newblock Semi-Supervised Classification with Graph Convolutional Networks.

\bibitem[{Li et~al.(2020)Li, Gupta, Morales, Allen, and
  Kochenderfer}]{li2020deep}
Li, S.; Gupta, J.~K.; Morales, P.; Allen, R.; and Kochenderfer, M.~J. 2020.
\newblock Deep Implicit Coordination Graphs for Multi-agent Reinforcement
  Learning.

\bibitem[{Oliehoek(2012)}]{Oliehoek2012}
Oliehoek, F.~A. 2012.
\newblock \emph{Decentralized POMDPs}, 471--503.
\newblock Berlin, Heidelberg: Springer Berlin Heidelberg.
\newblock ISBN 978-3-642-27645-3.
\newblock \doi{10.1007/978-3-642-27645-3_15}.
\newblock \urlprefix\url{https://doi.org/10.1007/978-3-642-27645-3_15}.

\bibitem[{Schulman et~al.(2017)Schulman, Wolski, Dhariwal, Radford, and
  Klimov}]{schulman2017proximal}
Schulman, J.; Wolski, F.; Dhariwal, P.; Radford, A.; and Klimov, O. 2017.
\newblock Proximal Policy Optimization Algorithms.

\bibitem[{Singh, Jain, and Sukhbaatar(2018)}]{singh2018learning}
Singh, A.; Jain, T.; and Sukhbaatar, S. 2018.
\newblock Learning when to Communicate at Scale in Multiagent Cooperative and
  Competitive Tasks.

\bibitem[{Sunehag et~al.(2017)Sunehag, Lever, Gruslys, Czarnecki, Zambaldi,
  Jaderberg, Lanctot, Sonnerat, Leibo, Tuyls, and
  Graepel}]{DBLP:journals/corr/SunehagLGCZJLSL17}
Sunehag, P.; Lever, G.; Gruslys, A.; Czarnecki, W.~M.; Zambaldi, V.~F.;
  Jaderberg, M.; Lanctot, M.; Sonnerat, N.; Leibo, J.~Z.; Tuyls, K.; and
  Graepel, T. 2017.
\newblock Value-Decomposition Networks For Cooperative Multi-Agent Learning.
\newblock \emph{CoRR} abs/1706.05296.
\newblock \urlprefix\url{http://arxiv.org/abs/1706.05296}.

\bibitem[{Veličković et~al.(2018)Veličković, Cucurull, Casanova, Romero,
  Liò, and Bengio}]{veli}
Veličković, P.; Cucurull, G.; Casanova, A.; Romero, A.; Liò, P.; and Bengio,
  Y. 2018.
\newblock Graph Attention Networks.

\end{thebibliography}
\end{document}